# Aggression in Hindi and English Speech: Acoustic Correlates and Automatic Identification

**Ritesh Kumar, Atul Kr. Ojha, Bornini Lahiri, Chingrimnng Lungleng**

Dr. Bhimrao Ambedkar University, Jawaharlal Nehru University, Indian Institute of Technology, Tezpur University

Agra, New Delhi, Kharagpur, Tezpur

E-mail: ritesh78_llh@jnu.ac.in, bornini@hss.iitkgp.ac.in, {shashwatup9k, chingriwo.lungleng}@gmail.com

**Abstract**

In the present paper, we will present the results of an acoustic analysis of political discourse in Hindi and discuss some of the conventionalised acoustic features of aggressive speech regularly employed by the speakers of Hindi and English. The study is based on a corpus of slightly over 10 hours of political discourse and includes debates on news channel and political speeches.
Using this study, we develop two automatic classification systems for identifying aggression in English and Hindi speech, based solely on an acoustic model. The Hindi classifier, trained using 50 hours of annotated speech, and English classifier, trained using 40 hours of annotated speech, achieve a respectable accuracy of over 73% and 66% respectively. In this paper, we discuss the development of this annotated dataset, the experiments for developing the classifier and discuss the errors that it makes.

**Keywords:** Aggression, Hindi, English, Acoustic Analysis, Classification, Computational Pragmatics.

## 1. Introduction

Verbal aggression could be understood as any kind of linguistic behaviour which intends to damage the social identity of the target person and lower their status and prestige (Baron and Richardson, 1994, cited in Culpeper, 2011). It can be expressed covertly (which is largely a result of unratified verbal behaviour in the given context but many not necessarily employ explicit, conventionalised linguistic structures) as well as overtly (which is largely expressed through conventionalised linguistic structures including prosody and syntactic structures). We would like to emphasise here that aggression is not an emotion (and so the task of detecting aggression is not a sub-task of emotion detection); rather it is a behaviour which may be the result of different kinds of emotions experienced by the speakers as well as it may result in different kinds of emotions in hearers. It is important to maintain this distinction as research and advancement in emotion distinction may actually feed into aggression detection and vice-versa and considering one a part of the other may hinder exploring this interaction.

In our present research, we largely look at overt aggression in political speeches and debates. It has been observed that covert aggression is relatively less frequent in these kinds of contexts and because of the lack of obvious cues, it is more difficult to automatically recognise it.

Overt aggression is, generally, accompanied by a number of cues like increase in the fundamental frequency of the speech, the amplitude, and the relative duration of the voiced part of speech and at the same time decrease in the importance of the unvoiced part and the spectral tilt. It also leads to the reduction of voice quality which is caused by a loss of control over the vocal folds. All these cues could be used for the detection of aggression in speech. These cues could be isolated from the speech using signal processing techniques. It has been reported by the previous experiments that best cues for aggression include fundamental frequency, the ratio of signal energy below and above 1000 Hz and the standard deviation of the energy of the three highest peaks in the spectrum. Some of the other acoustic cues that could be used for detection of aggression include level cue, audibility cue, spectrum distortion cue, harmonic distortion cue, pitch salience cue and pitch height cue (van Hengel and Andringa, 2007).

Covert aggression, on the other hand, lacks cues in the speech itself. However, it could be inferred based on the response of the hearer. It has been argued in the theoretical aggression literature that silence on the part of the hearer in certain circumstances (for example, as response to a compliment) could be indicative of covert aggression.

As we will see in section 4, we have annotated all kinds of aggression in the corpus and the final training of the recognition system needs to take both the acoustic features of the annotated speech as well as these theoretical insights into consideration. However, since the data that we use for training is mostly news interviews and not actual conversation, most of these features do not play a significant role in the current experiments.

## 2. Corpus collection and annotation

The present corpus is being developed with a specific purpose in mind – automatic detection of aggressive behaviour in speech using. So most of the design and compilation decisions are taken keeping this goal in mind.

Currently the corpus consists of data from the following sources -

1. **Political Debates on News Channels** – These are the panel discussions and debates aired on some of the major news channels. The debates included in the corpus are taken from those available in the public domain over the web. These consist of around 60 hours of total data in English and Hindi each. The data was downloaded either from YouTube or the official websites of the news channels using the VideoDownloadHelper plugin and saved in the best available quality in MP4 format.

2. **Political Speeches** – In addition to the political debates, political speeches available on the web are also included in the corpus. However, unlike the debates these are monologues and are largely aggressive. These make up around 80 hours of total data in Hindi and English each. These speeches were downloaded using the same methods as above and similar standards were followed for their storage.

In addition to this, audio of the complete data was also extracted in .wav format and stored separately as the speech corpus. It was done for two reasons – it is comparatively easier to store, maintain, process and share speech data than multimodal data and more importantly, for the current purposes, we only needed speech data as we would be training the system based only on speech (more specifically prosodic) and textual features. Moreover, it would not be a mammoth task to integrate the video with this annotated and transcribed speech data at a later stage if we need to use the video features as well.

This speech corpus is transcribed using Praat at two tiers - Aggression and Turn taking – which is motivated by the empirical findings that aggression can be understood and recognised very significantly by referring to the conversation structures, especially turn-taking and preference organisation in the discourse. On tier 1, the level and kind of aggression was annotated. It is to be noted that there is hardly any objective measure of aggression and the annotation of aggression level was largely carried out on the basis of the annotator's own impression of the speech.

On Tier 2, the turns in a conversation are annotated. These tags are motivated by the concepts regularly employed in Conversation Analysis (CA) – turn construction unit (TCU), overlap and interruption. These 3 concepts are central to the understanding of the organisation and structure of turns in a conversation. These concepts are employed for annotation here so that turns may be adequately classified and also correlated with different kinds of aggression. A brief description of each tag on the two tiers is given in the following sections.

### 2.1 Overtly Aggressive Threatening (OAG_T)

Any segment of speech in which aggression is overtly expressed – either through the use of a specific kind of prosody which 'sounds' aggressive and / or certain lexical / syntactic structures – and which may lead to some kind of conflict or physical violence is to be annotated using this label. One possible instance will be aggressive exchange on the road, which has the possibility of turning ugly.

### 2.2 Overtly Aggressive Non-Threatening (OAG_NT)

Any segment of speech in which aggression is overtly expressed – either through the use of specific kind of prosody which 'sounds' aggressive and / or certain lexical / syntactic structures – but which is highly unlikely to lead to some kind of conflict or physical violence is to be annotated using this label. One possible instance will be aggressive discussion on the TV channel or in public forums where the speakers are not likely to engage in physical aggression.

### 2.3 Covertly Aggressive Threatening (CAG_T)

Any segment of speech in which aggression is not overtly expressed but which may still lead to some kind of conflict or physical violence is to be annotated using this label. One of the possible cases would be eve-teasing instances which may lead to more serious offense like physical assault and rape.

### 2.4 Covertly Aggressive Non-Threatening (CAG_NT)

Any segment of speech in which aggression is neither overtly expressed nor does it have possibility to lead to some kind of conflict or physical violence is to be annotated using this label. One of the possible cases would be gendered, racial or casteist speeches in public forums which is highly unlikely to lead to actual conflict and violence by the speakers themselves and may not be spoken using conventionalised aggressive structures.

### 2.5 Non-aggressive
This label should be given to all those human speech samples which do not exhibit aggression in any form.

### 2.6 Irrelevant
This label should be given to only those samples which do not contain any kind of human speech at all. Those portions of the speech which are blank or which contain noise or animal sounds are to be given this label. However, it should be kept in mind that slight pauses (of say, upto 5 seconds) should not be separately marked by this label). It is to be used when it is evident that there is no pause intended and the recording is not part of a human speech. Generally noise/music/silence at the beginning or towards the end of a speech file is to be marked by this label such that the beginning and end of an aggression or non-aggressive human speech is clearly marked out.

### 2.7 Turn Construction Unit (TCU)
It is a largely accepted fact that we don't always talk in sentences: a single word, a clause, a phrase, a sentence, or even a gesture can count as a meaningful and complete contribution. Each of these construct, at the end of which (called transition-relevance place (TRP) in the CA literature), there is a possibility of the completion of turn, is called Turn Construction Unit. TRP is where turn-taking can occur without affecting the speaker's turn and the construct in between two TRPs or in between the beginning of a turn and a TRP is the TCU. Each turn of a speaker may contain several smaller TCUs and at the same time the complete turn, if not interrupted, is a complete TCU in itself. For the present purposes, we are annotating only the complete TCU as a TCU and not the smaller TCUs within it. Thus effectively we mark out each turn in the conversation using this tag.

### 2.8 Overlap
During a conversation when a speaker, other than the present speaker, takes turn and starts speaking just when TCU is about to end such that for a few milliseconds both the speakers are speaking at the same time then the whole period for which both are speaking is to be marked as overlap.

### 2.9 Interruption
Interruption refers to the act of taking turns when one speaker has not yet finished speaking and there is no TRP (end of TCU) in the next few milliseconds and the other speaker starts speaking. In order to distinguish between overlap and interruption the annotators will have to make a judgement as to whether TRP was very close when the other speaker took turn or it was not possible to guess the TRP when interruption occurred. Generally, in overlap, the first speaker finishes her/his turn with TCU during the period when both speakers are speaking but in interruption, there is no proper TRP reached in the first speaker's speech as the second speaker completely takes over the turn.

Using this annotation scheme, we annotated a total of 50 hours of speech / conversation data in Hindi and 40 hours in English. We use this annotated dataset for the automatic classification experiments as well as theoretical study discussed below.

## 3. Acoustic Correlates of Aggression
We carried out a corpus-based study of acoustic / prosodic features of aggression in both English and Hindi. The study was carried out using the annotated dataset of approximately 10 hours of speech / conversation in each of the languages. We used 3 categories – Covertly Aggressive (CAG), Overtly Aggressive (OAG) and Non-aggressive (NAG) – for the study.

We looked at a set of 9 acoustic features in each of the approximately 1500 speech samples in each of the 2 language. The distribution of each category in terms of number of speech samples in each language is summarised in Table 1 below.

| Category | Hindi | English |
|---|---|---|
| Covertly Aggressive | 245 samples | 260 samples |
| Overtly Aggressive | 618 samples | 620 samples |
| Non-aggressive | 709 samples | 697 samples |
| Total | 1572 samples | 1577 samples |

Table 1: Count of speech samples used for study

The features selected for the study are taken from the Geneva Minimalistic Acoustic Parameter Set (GeMAPS) that was devised for conducting research on emotion, voice and affective computing [4]. The features included in the GeMAPS is grounded in a large number of previous research in psychology as well as computational linguistics (see Eyben et al., 2016; Scherer, 2003 and Goudbeck and Scherer, 2010 for a review of such studies). These features are included on the basis of their potential to index physiological changes in voice production during affective processes and also on the basis of the empirical evidence of their adequacy in previous research. The acoustic features tested in the study are as below -

- **F0** - it roughly corresponds to Pitch
- **Intensity** - it roughly corresponds to loudness of the voice
- **Jitter** - Deviations in individual consecutive F0 period lengths
- **Shimmer** - Difference of the peak amplitudes of consecutive F0 periods
- **Rate of Loudness Peaks (RLP)** - Number of loudness peaks per second
- **Mean length of voiced regions** (MVd)
- **Mean length of voiceless regions** (MVl)
- **Number of continuous voiced regions per second** (CVd)
- **Spectral Flux** (SF) - Difference of the spectra of two consecutive frames

A comparison of the means of these different features is given in Figure 1.

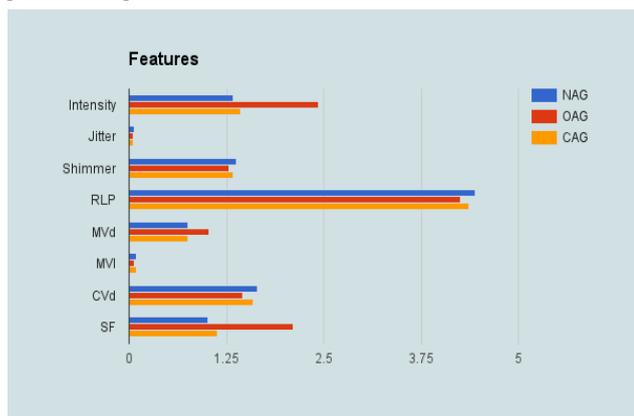

Figure 1: Mean of different features in different categories in Hindi

An analysis of variance (ANOVA) was carried out to test if the difference between the mean value of each of these features in each category is significantly different or not. The results showed that every feature under study has significantly different means for the three categories i.e. all of these acoustic features are correlated with aggression (p = 0 for all the features).

However, further Tukey pairwise tests (at 0.01 and 0.05 levels of significance) gave a much more nuanced view of the interrelation between acoustic features and aggression. The only feature that has distinct, statistically significant mean value for all the three categories of overtly aggressive, covertly aggressive and non-aggressive is shimmer.

The features that have significantly different values in overtly and covertly aggressive speech are -

- Intensity/Loudness
- Spectral Flux
- Mean length of Voiced and Voiceless regions
- Jitter (to some extent, only at 0.05 level of significance)

Similar analysis shows that covertly aggressive and non-aggressive speech could be distinguished only in their use of following feature -

- F0/Pitch
- Jitter (to some extent, only at 0.05 level of significance)

However, not very surprisingly but yet quite unexpectedly, in cases of non-aggressive and overtly aggressive speech, each of 9 features play a distinguishing role.

What is even more significant is the fact that these results are exactly the same for both English as well as Hindi i.e. for both the languages shimmer has statistically distinct mean value for all the three different categories. Moreover, overtly and covertly aggressive Hindi speech is distinguished by differences in loudness, spectral flux, mean length of voiced and voiceless regions and jitter. Similarly, covertly aggressive and non-aggressive speech in Hindi is distinguished by Pitch and Jitter while overtly aggressive and non-aggressive speech is distinguished by all the 9 features discussed above. Based on the results of this study, it could be postulated that, at least, some of these acoustic features might be indexical of aggressive speech. As such these features may be employed for automatic detection of aggressive / dangerous speech.

In the next section, we discuss the development of an automatic classification system based on the findings of this study.

## 4. Classification Experiments and Results

We used a total dataset of approximately 50 hours in Hindi and 40 hours in English to train two classifiers – one each for English and Hindi – for automatic identification of aggression in speech. The distribution of each class in the dataset is given in Table 2 below

| Category | Hindi | English |
|---|---|---|
| **Overtly Aggressive** | 2240 samples | 3761 samples |
| **Covertly Aggressive** | 347 samples | 615 samples |
| **Non-aggressive** | 5370 samples | 4618 samples |
| **Total** | 7957 samples | 8994 samples |

Table 2: Count of speech samples in the dataset

As one would notice, even if the quantity of English data in terms of the duration is less than that of Hindi data, the number of samples is greater than Hindi. This is because of the nature of the data. The samples are not of equal length and are marked by the annotators based on their perception. So if there is a large continuous region that is marked as of one category, it is taken as one sample. As such, it could be assumed that English data shows more variation in terms of aggression and so there are smaller duration of aggressive / non-aggressive speech.

For training, acoustic features[1] were extracted using openSMILE (Eyben et al., 2013), an open source toolkit for speech processing. The dataset was divided into 80:10:10 train:validate:test sets. The dataset was used to train an SVM multiclass classifier. The parameters for the classifier were optimised using a Grid Search approach.

### 4.1 Feature Ablation Experiments

Based on the study conducted in the previous section, we conducted a feature ablation study to understand the actual role of each feature in the classification process. We conducted a set of 3 experiments and looked at how each of the findings of the study hold in each of these experiments. These experiments are as discussed below:

a. **Experiment 1**: In the first experiment, we used only 'shimmer' as the feature for experiments. We used both the functionals – mean and standard deviation - for shimmer, provided by the OpenSmile toolkit, as features for the experiments. Since shimmer came out to be the distinguishing feature across all the three classes, we wanted to check how effective it would be for classification purposes.

b. **Experiment 2**: In the second set of experiments, we added pitch (since that distinguished between covert aggression and non-aggression) and then jitter. After that we added the 4 other features i.e. intensity, spectral flux and mean length of voiced and voiceless regions that distinguished between covert and overt aggression one-by-one and finally we added the remaining 2 other features used in the study i.e. rate of loudness peaks and number of continuous voiced regions per second. For pitch and intensity, there are 10 functionals included in OpenSmile – we used all these functionals for the experiments. Similarly, we used the 2 functionals each for jitter and spectral flux. Taken together, these made up a total of 28 features.

c. **Experiment 3**: In the final experiments we added the remaining 17 low-level descriptors, making up a total of 60 features, to see if they have some effect on the performance of the classifiers. These included MFCC features, F1, F2, F3 related features and other signal-related features

The results of this feature ablation study is summarised in Table 3.

| Feature Set | Accuracy | |
|---|---|---|
| | Hindi | English |
| Shimmer | 67.71 | 51.89 |
| +F0 | 67.71 | 57.56 |
| +Jitter | 67.71 | 58.0 |
| +Intensity | 67.71 | 63.67 |
| +Spectral Flux | 67.84 | 63.67 |
| +Mean length of voiced and voiceless regions | 67.84 | 64.33 |
| +Rate of loudness peaks + number of continuous voiced regions per second | 67.84 | 64.78 |
| + remaining features included in GeMAPS viz. MFCC, F1, F2, F3, etc. | 73.37 | 66.33 |

Table 3: Results of the feature ablation study

### 4.2 Results and Discussion

The best classifier gave a score of over 66% and 73% respectively for English and Hindi on the held-out test set using all the 88 features. The F1 scores, however, are lower at 0.64 and 0.68 respectively for English and Hindi. The test set contained 900 samples of English and 796 samples of Hindi. The classification report for both the languages is given in Table 4. As it is clear, for Hindi, while precision is acceptable for OAG, recall is quite low, thereby, significantly lowering the overall performance of the classifier.

| Metric | Language | OAG | CAG | NAG |
|---|---|---|---|---|
| Precision | Hindi | 0.71 | 0.0 | 0.74 |
| | English | 0.65 | 0.0 | 0.68 |
| Recall | Hindi | 0.30 | 0.0 | 0.96 |
| | English | 0.67 | 0.0 | 0.74 |
| F1 | Hindi | 0.42 | 0.0 | 0.83 |

---
[1] The acoustic features included the 9 features discussed in the study discussed in the previous section as well as 79 additional features (making it a total of 88 features) included in the GeMAPS feature set.

|  | English | 0.66 | 0.0 | 0.71 |
|---|---|---|---|---|
| **Overall Scores** | | | | |
|  | Precision | Recall | F1 | |
| Hindi | 0.70 | 0.73 | 0.68 | |
| English | 0.62 | 0.66 | 0.64 | |

Table 4: Classification Report

One of the most interesting takeaways of this study was the difference in how the classifiers for Hindi and English differed with respect to different features. While for Hindi, shimmer seems to be the most distinguishing feature, for English, almost every feature seems to be relevant. The performance of Hindi classifiers improved marginally with addition of spectral flux and it improves by approximately 5% only when the energy and signal-related features are utilised. On the other hand, for English, barring spectral flux, addition of every other feature led to an improvement in the performance. Thus even though our earlier study showed that similar features would work for both the languages, it does not seem to be working in practice and needs to be further investigated.

Table 5 and 6 below show the confusion matrix of the best classifier for English and Hindi data respectively.

| Category | OAG | CAG | NAG | Total |
|---|---|---|---|---|
| **OAG** | 252 | 0 | **123** | 375 |
| **CAG** | 15 | 0 | **43** | 58 |
| **NAG** | **121** | 1 | 345 | 467 |

Table 5: Confusion Matrix for English Classifier

| Category | OAG | CAG | NAG | Total |
|---|---|---|---|---|
| **OAG** | 67 | 0 | **155** | 222 |
| **CAG** | 5 | 0 | **30** | 35 |
| **NAG** | **22** | 0 | 517 | 539 |

Table 6: Confusion Matrix for Hindi Classifier

The classification results show that the classifier is largely confused in covert aggression and it is marked as both OAG as well as NAG. This clearly shows that the classifier has not been able to learn CAG. This could be because of the smaller number of samples for CAG in the training set and also because of the presence of clearly discriminating acoustic features for this class. As expected, the system performs best in recognising overt aggression in the speech – besides the presence of obvious acoustic cues, it could also be because of a relatively large number of samples available for training. The classifier is also able to classify NAG with a fair amount of accuracy. However, what is surprising is that there are quite a few instances where NAG is classified as OAG and vice-versa. A closer look at the misclassified samples reveal that length of the sample could be one of the reasons for these misclassification. Generally, NAG samples are much bigger than OAG samples in terms of their length and as such a lot of larger samples are marked as NAG. Moreover, some of the misclassified samples are quite small and so, probably, does not provide sufficient features for the classifier to classify it appropriately.

## 5. Summing Up

In this paper we have discussed a corpus-based study of acoustic correlates of aggression in Hindi and English speech / conversation. The study showed that some of the acoustic / prosodic features like loudness, jitter, shimmer, pitch, etc. are very strongly correlated with covert / overt aggression cross-linguistically and so these could prove to be very helpful in automatic identification of aggression in speech. On the basis of this study, we trained 2 classifiers – one each for English and Hindi - that could automatically identify aggression in speech. The classifiers were trained using only 88 GeMAPS acoustic features (without the use of any language model) and the classifiers for English and Hindi gave a score of over 66% and 73% respectively. This score could be further improved with the use of a more balanced dataset and more samples for CAG. Additionally we need to experiment with language models also, which might be helpful in further improvement of the system.

## Acknowledgments

This research was supported by UGC-UK-India Education and Research Initiative grant and was carried out jointly in collaboration with Prof. Daniel Kadar of University of Huddersfield, UK and his team.

## 6. Bibliographical References

Baron, R.A. and D.R. Richardson, 1994, *Human Aggression,* New York: Plenum Press.

Culpeper, J, 2011, *Impoliteness Using Languages to cause Offence*. Cambridge: Cambridge University Press

Van Hengel, P. W. J. and T. C. Andringa, 2007, Verbal aggression detection in complex social environments, *AVSS, IEEE Computer Society*, pp. 15-20

Eyben, F., Klaus R. Scherer, Bjorn W. Schuller, Johan Sundberg, Elisabeth Andre, Carlos Busso, Laurence Y. Devillers, Julien Epps, Petri Laukka, Shrikanth S. Narayanan, and Khiet P. Truong, 2016, The Geneva Minimalistic Acoustic Parameter Set (GeMAPS) for Voice Research and Affective Computing, *IEEE Transactions on Affective Computing,* Vol. 7, No. 2, pp. 190 – 202

Scherer, K. R., 2003, Vocal communication of emotion: A review of research paradigms, *Speech Communication*, vol. 40, pp. 227–256.

Goudbeek, M. and K. R. Scherer, 2010, Beyond arousal: Valence and potency/control cues in the vocal expression of emotion, *Journal of the Acoustical Society of America (JASA)*, vol. 128, pp. 1322–1336.

Eyben, F., F. Weninger, F. Gross, and B. Schuller, 2013, Recent developments in openSMILE, the munich open-source multimedia feature extractor, in *Proc. of ACM MM 2013, Barcelona, Spain*. New York, NY, USA: ACM, pp. 835–838.